\documentclass[PRD,reprint,twocolumn,linenumbers,amssymb,amsmath,aps,showpacs,nofootinbib,superscriptaddress]{revtex4}
\usepackage{graphicx}
\usepackage{dcolumn}
\usepackage{bm}
\usepackage{color,soul}
\setulcolor{red}

\newcommand{\beq}{\begin{equation}}
\newcommand{\eeq}{\end{equation}}
\newcommand{\beqa}{\begin{eqnarray}}
\newcommand{\eeqa}{\end{eqnarray}}

\begin{document}

\title{Light bending in infinite derivative theories of gravity}

\author{Lei Feng}
\email{fenglei@pmo.ac.cn}
\affiliation{Key Laboratory of Dark Matter and Space Astronomy, Purple Mountain Observatory, Chinese Academy of Sciences, Nanjing 210008, China}
\affiliation{Joint Center for Particle, Nuclear Physics and Cosmology, Nanjing 210093, China}

\begin{abstract}
 Light bending is one of the significant predictions of general relativity (GR) and it has been confirmed with great accuracy during the past one hundred years. In this paper, we semiclassically calculate the deflection angle for the photons that just grazing the Sun in the infinite derivative theories of gravity (IDG) which is a ghost and singularity free theory of gravity.  From our calculations, we find that the deflection angle $\theta$ only depends on $\Lambda/E$. $\theta \rightarrow \theta_E$ when $\Lambda/E \rightarrow \infty$ and decrease to zero when $\Lambda/E \rightarrow 0$. The transition interval occurs at $10^{4}< E/\Lambda < 10^{7}$. It should be pointed out that this model can be tested by the Chandra X-ray Observatory if $0.01 eV < \Lambda < 0.1 eV$.
\end{abstract}
\maketitle

\section{Introduction}

The general relativity (GR) achieved great success in the past one hundred years and has been tested through different kinds of experiments. Light bending is one important test which was first observed by Eddington and Dyson in 1919. Many measurements were made in the following years and the accuracy is greatly improved using the very long baseline radio interferometry\cite{lebach,beson}. The GR theory fits the experiments very well so far.

Unfortunately, the quantum GR is not perturbatively renormalizable. The higher-derivative gravity (HDG) theory could avoid such difficulties. HDG was first introduced by Weyl \cite{Weyl} and Eddington \cite{Eddington} which includes the higher-derivative terms in the Lagrangian such as scalars $R^2, R^2_{\mu \nu},$, $R^2_{\mu \nu \alpha \beta}$ and so on.
Such models are renormalizable\cite{Stelle} but nonunitary at the same time and
and it is unavoidable that the ghost particles emerge when the higher derivatives are introduced.
The infinite derivative theories of gravity (IDG)\cite{IDG1,IDG2,IDG3} is such a model that can avoid the problem of missive ghost(It should be noted that this model is also named super-renormalizable quantum gravity or super-renormalizable nonlocal quantum gravity in some other papers). More details of IDG can be found in\cite{buoninfante}. An earlier similar theory can also be found in \cite{Biswas}.

The significant advantages of IDG are that it could avoid the problem of massive spin-2 ghost and the divergence of gravitational potential at small distance. The gravitational action for IDG can be written as\cite{IDG1,IDG2,IDG3}

\begin{equation}
S=-\frac{1}{16\pi G}\int d^4x \sqrt{-g} \{R+ G_{\mu\nu}\frac{a(\square)-1}{\square}R^{\mu\nu}\},
\label{eq:s}
\end{equation}
where $\square$ is the D'Alambertian operator, $G_{\mu\nu}=R_{\mu\nu}-\frac{1}{2}g_{\mu\nu}R$ and $a(\square)=e^{-\frac{\square}{\Lambda^2}}$. It should be noted that $\Lambda$ corresponds to a non-locality scale because the gravitational interactions in IDG model is non-local.
The lower limit on parameter $\Lambda$ can be calculated by combining the result of \cite{lowerlimit,lowerlimit2}, which is
\begin{equation}
\Lambda > 0.01~eV.
\end{equation}
It should noted that the authors studied the much more general situation($a(\Box) = exp[(\Box/\Lambda^2)n]$) in\cite{lowerlimit}, and got much stronger lower bound on $\Lambda$ for higher n. In \cite{Leandros}, the author found that there was evidence for a Newtonian potential with of the form in\cite{lowerlimit}. Much stronger bound on $\Lambda$ was gotten by using IDG as an extension of Starobinsky inflation\cite{Edholm}.

In\cite{HDG1,HDG2,HDG3}, the authors studied the gravitational deflection within the framework of classical and semiclassical HDG and find that the deflection angle decreases to zero at $log_{10}|\beta|\sim 89$ in classical HDG and $log_{10}|\beta|\sim 70$ in semiclassical HDG. The deflection angel calculated in other gravitational theory can be found in \cite{Accioly1,Accioly2,Accioly3,Accioly4,Accioly5,Hu}. In this paper, we calculate the gravitational deflection of photons that graze the sun within the framework of semiclassical IDG.

This draft is organized as follows: In Sec. II we calculate the deflection angle in semiclassical IDG and our conclusions are summarized in Section III. Here we use natural units and diag(1, -1, -1, -1) as the Minkowski metric.

\section{Gravitational deflection in tree-level IDG}
We solve the linearized field equations of IDG for a pointlike particle using the perturbed metric
\begin{equation}
g_{\mu \nu}= \eta_{\mu \nu} + \kappa h_{\mu \nu},
\end{equation}
where $\kappa=\sqrt{16\pi G}$.
The field equations derived from the action in Eq.(\ref{eq:s}) with a source component is\cite{buoninfante}
\begin{eqnarray}
a(\square)[\square h_{\mu \nu} -(\partial_{\mu}\partial_{\alpha}h_{\nu}^\alpha+\partial_{\alpha}\partial_{\nu}h_{\mu}^\alpha)+ \\ \nonumber
(\eta_{\mu \nu}\partial_{\alpha}\partial_{\beta}h^{\alpha\beta}+\partial_{\mu}\partial_{\nu}h)-\eta_{\mu \nu}\square h]=-\kappa T_{\mu \nu},
\end{eqnarray}
where $T_{\mu \nu}$ is the energy-momentum tensor of the source term. The corresponding energy momentum tensor for a particle with mass M is
$ M\eta_{\mu 0} \eta_{\nu 0}\delta^3({\bf r})$. Solving the above equation with such energy momentum tensor\cite{IDG1,IDG2,IDG3,buoninfante}
 and we find
\begin{equation}
h_{\mu \nu}(r)= \frac{M\kappa}{8 \pi} \Big[ \frac{\eta_{\mu \nu}}{r} - \frac{2\eta_{\mu 0} \eta_{\nu 0}}{r}\Big]Erf(\frac{\Lambda r}{2}).
\label{eq:r}
\end{equation}

In this model, the modified Newtonian potential is
\begin{equation}
\phi(r)=\frac{\kappa}{2}h_{00}=-\frac{GM}{r}Erf(\frac{\Lambda r}{2}).
\end{equation}
It should be pointed out that such kind of potential was first obtained by Tseytlin in the exponential gravity motivated by string theory\cite{Tseytlin}. $Erf(\frac{\Lambda r}{2})\rightarrow 0$ when $r\rightarrow \infty$ and we recover the Newtonian potential. $Erf(x)\sim \frac{2}{\sqrt{\pi}}e^{-x^2}x \sim\frac{2}{\sqrt{\pi}}x$ when $x\rightarrow 0$ and then
\begin{equation}
\phi(r)\rightarrow -\frac{GM\Lambda}{\sqrt{\pi}},
\end{equation}
So IDG model can avoid the divergence problem in GR at small distance.

\begin{figure}
\includegraphics[width=85mm,angle=0]{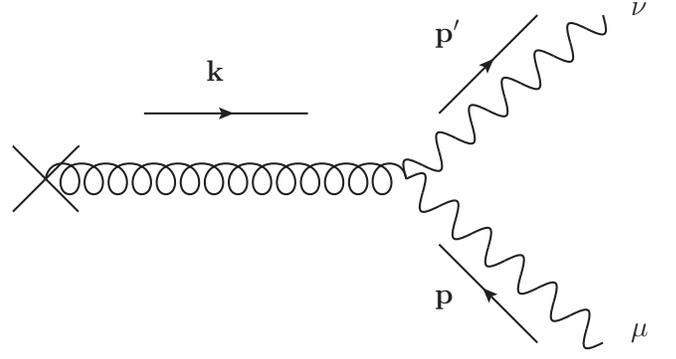}
  \caption{The Feynmann diagram of the interaction between external gravitational field and photon.}
  \label{fig:1}
\end{figure}

Similar with\cite{HDG1,HDG2,HDG3,Accioly1,Accioly2,Accioly3,Accioly4,Accioly5}, we calculate the gravitational deflection angle within the framework of semiclassical IDG. This method provides much more information about the gravitational deflection of photons, such as the energy dependence of the deflection angle and so on. The Feynman diagram of this process that the photon  scattered by the external gravity field is shown in Fig.\ref{fig:1} and the corresponding amplitude  is given by

\begin{eqnarray}
{\cal{M}}_{r r'} =&& \frac{1}{2} \kappa h^{\lambda \rho}_{\mathrm{ext}}({\bf{k}})\Bigg[-\eta_{\mu \nu} \eta_{\lambda \rho}pp' + \eta_{\lambda \rho}p'_\mu p_\nu + 2\Big(\eta_{\mu \nu}p_\lambda p'_\rho  \nonumber \\  &&- \eta_{\nu \rho}p_\lambda p'_\mu  \nonumber - \eta_{\mu \lambda }p_\nu p'_\rho + \eta_{\mu \lambda} \eta_{\nu \rho}pp'\Big) \Bigg]\epsilon^\mu_r({\bf{p}}) \epsilon^\nu_{r'}({\bf{p'}}) \nonumber,
\end{eqnarray}

\noindent where $\epsilon^\mu_r({\bf{p}})$ ($\epsilon^\nu_{r'}({\bf{p'}})$) denotes the polarization vectors of the initial (final) photons and  satisfies the following relation

\begin{equation}
\sum_{r=1}^{2}\epsilon^\mu_r({\bf{p}})\epsilon^\nu_{r}({\bf{p}})= -\eta^{\mu \nu} - \frac{p^\mu p^\nu}{(p\cdot n)^2} + \frac{p^\mu n^\nu + p^\nu n^\mu}{p\cdot n},
\end{equation}

\noindent where $n^2=1$. Here $h^{\lambda \rho}_{\mathrm{ext}}({\bf{k}})$ is the gravitational field in momentum space, which is

\begin{equation}
h^{\lambda \rho}_{\mathrm{ext}}({\bf{k}})= \int{d^3{\bf{r}} e^{-i{\bf{k}}\cdot {\bf{r}}}h^{\lambda \rho}_{\mathrm{ext}}({\bf{r}})}.
\label{eq:k}
\end{equation}

\noindent Substituting Eq.(\ref{eq:r}) into Eq.(\ref{eq:k}), we get
\begin{equation}
h^{{\mathrm{(E)}} \mu \nu}_{\mathrm{ext}}({\bf{k}}) = \kappa  M\left(\frac{\eta^{\mu \nu}}{2{\bf{k}}^2} - \frac{ \eta^{\mu 0} \eta^{\nu 0}}{\bf{\bf{k}}^2}\right){\rm exp(-\frac{{\bf{k}}^2}{\Lambda^2}}).
\end{equation}

Then we get the unpolarized cross-section with the following equation
\begin{eqnarray}
\frac{d\sigma}{d \Omega}&&= \frac{1}{(4 \pi)^2} \frac{1}{2}\sum_{r} \sum_{r'}{\cal{M}}^2_{r r'} \nonumber \\ &&= \frac{1}{(4 \pi)^2 }\frac{\kappa^4 M^2 E^4(1 + \cos{\theta})^2}{4} \Bigg[\frac{1}{{\bf{k}}^2}{\rm exp(-\frac{{\bf{k}}^2}{\Lambda^2})} \Bigg]^2, \nonumber
\end{eqnarray}
where $\theta $ is the angle between $\bf p$ and $\bf p^\prime$ and $E$ is the energy of the injected photon.

For small angles case, $\rm {\bf k^2}\approx 2{\bf p^2}(1-cos\theta)\approx E^2\theta^2$. Then the previous equation reduces to
\begin{eqnarray}
\frac{d\sigma}{d \Omega}= 16 G^2 M^2 \Big[\frac{1}{\theta^2}{\rm exp(-\frac{\theta^2}{\lambda^2})} \Big]^2.
\end{eqnarray}
where $\lambda \equiv \frac{\Lambda}{E}$. Obviously, the cross section only depend on $\lambda$.

From the above equation, we can see that
\begin{eqnarray}
\frac{d \sigma}{d \Omega} \rightarrow  \Big(\frac{4GM}{\theta^2} \Big)^2,\;\; {\mathrm{if}} \; \lambda \rightarrow \infty;
\end{eqnarray}
i.e. we recover the standard cross section of GR and the deflection angle is $1.75^{\prime\prime}$.
And
\begin{eqnarray}
\frac{d \sigma}{d \Omega} \rightarrow 0, \;\; {\mathrm{if}} \; \lambda \rightarrow 0,
\end{eqnarray}
which means that the deflection angle decreases to zero when $\lambda \rightarrow  0$.

We compare the classical and the tree-level cross-section formulas to get the classical particle trajectory\cite{Delbourgo,Berends}

\begin{eqnarray}
\frac{d\sigma}{d \Omega}= 16 G^2 M^2 \Bigg[\frac{1}{\theta^2}{\rm exp(-\frac{\theta^2}{\lambda^2})} \Bigg]^2= -\frac{rdr}{\theta d\theta}.
\label{eq:j}
\end{eqnarray}

\begin{figure}
\includegraphics[width=85mm,angle=0]{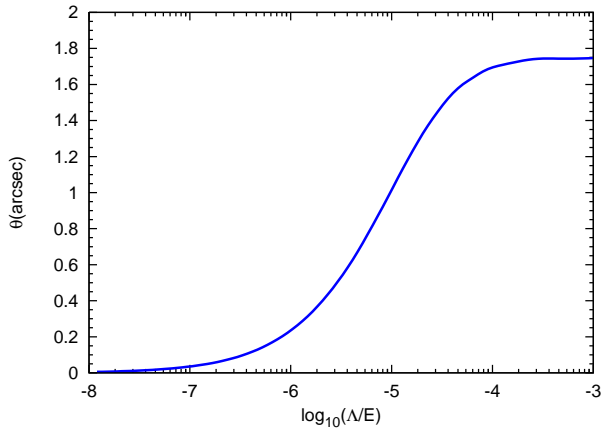}
  \caption{The deflection angle as a function of $\rm log_{10}(\Lambda/E)$ for photons that just grazing the Sun in semi-classical IDG.}
  \label{fig:2}
\end{figure}

Performing the integration on Eq.(\ref{eq:j}), we finally get the deflection angle for the photons that just grazing the Sun, which is

\begin{eqnarray}
\frac{1}{\theta^2_{\mathrm{E}}}= \frac{1}{\theta^2} exp(-\frac{2\theta^2}{\lambda^2}) +  \frac{2}{\lambda^2}Ei(-\frac{2\theta^2}{\lambda^2}),
\label{eq:a}
\end{eqnarray}
where $\theta_{\mathrm{E}}=\sqrt{4GM/R_{\bigodot}}=1.75^{\prime\prime}$ is the Einstein's deflection angle and $R_{\bigodot}$ is the radius of the sun. The exponential integral $Ei(x)$ is defined as
\begin{eqnarray}
Ei(x)=-\int_{-x}^{\infty}\frac{e^{-t}}{t}dt.
\end{eqnarray}
Defining $y\equiv \frac{2\theta^2}{\lambda^2}$ and Eq.(\ref{eq:a}) becomes
\begin{eqnarray}
\frac{e^{-y}}{y}+Ei(y)-\frac{\lambda^2}{2\theta_{\mathrm{E}}^2}=0.
\end{eqnarray}
We solve the above equation numerically and the result is shown in Fig.\ref{fig:2}.

\begin{table}[!htb]
\begin{center}
\caption {The performance of the current(or planed) X-ray(or $\gamma$-ray) detectors.}
\begin{tabular}{ccc}
\hline \hline
Detectors & Energy Range & Angular Resolution  \\
  \hline
Chandra\cite{chandra} & $\rm 0.1 - 10~keV$ & 0.5$^{\prime\prime}$ \\
Swift-XRT\cite{Swift} & $\rm 0.2 - 10~keV$ & 18$^{\prime\prime} {\rm HPD@1.5keV}$ \\
Swift-BAT\cite{Swift} & $\rm 15 - 150~keV$ & 17$^{\prime\prime}$ \\
FOXSI\cite{foxsi} & $\rm 5 - 15~keV$ & 12$^{\prime\prime}$ \\
XMM-Newton\cite{xmmnewton} & $\rm 0.1 - 12~keV$ &  $5^{\prime\prime} \sim 14^{\prime\prime}$ \\
NuSTAR\cite{NuSTAR} & $\rm 3 - 79~keV$ & 9.5$^{\prime\prime}$ \\
HXMT\cite{hxmt}& $\rm 1 - 250~keV$ & $<5^{\prime}$\\
Einstein Probe\cite{ep}& $\rm 0.5 - 4~keV$ & $<5^{\prime}$\\
Athena\cite{Athena} & $\rm 0.3 - 12~keV$ & 10$^{\prime\prime}$ \\
Fermi-LAT\cite{fermi} &    $\rm 10 - 3\times 10^5~MeV$  & $\rm \sim 0.1^\circ$ \\
DAMPE \cite{dampe} &    $\rm 5 - 10^4~GeV$  & $\rm \sim 0.1^\circ$ \\
CALET \cite{calet} &    $\rm 5 - 10^4~GeV$  & $\rm \sim 0.1^\circ$ \\
  \hline
  \hline
\end{tabular}
\label{table1}
\end{center}
\end{table}

From Fig.\ref{fig:2}, we can straightforwardly see that $\theta \sim \theta_{\mathrm{E}}$ as $E < 10^{4}\Lambda$, which recovers the result of GR and $\theta \sim 0$ as $E > 10^{7}\Lambda$ which means there is no deflection for sufficient high energy photons. It should point out that the transition interval occurs for $10^{4}\Lambda< E < 10^{7}\Lambda$. The smallest energy to test such effect is $10^{4}\Lambda$ and the deflection angle decrease to $1^{\prime\prime}$ at $E = 10^{5}\Lambda$.
If $0.01 eV<\Lambda<1 eV$, the transition occurs in X-ray band which could be tested by the X-ray telescopes. And the transition occurs in hard X-ray or gamma-ray band if $1 eV<\Lambda$ which could be tested by the corresponding telescopes.

The angular resolution should be better than $1.75^{\prime\prime}$ to test the deviation of deflection angle from $\theta_E$.
The performance of the current(or planed) X-ray(or $\gamma$-ray) detectors is shown in Table.\ref{table1}. From Table.\ref{table1}, we can see that only Chandra X-ray Observatory~(Chandra)\cite{chandra}~satisfies such requirement. Chandra works in the photon energy range of 0.2-10 keV. So IDG model can be tested if $0.01 eV < \Lambda < 0.1 eV$.
It is a big challenge to avoid the damage of detectors when doing such measurements because the sun is the brightest X-ray source in the sky. A large part of the X-ray are sheltered by the moon when the total solar eclipse occurs. So it may be possible to measure the X-ray deflection angle during the total solar eclipse with Chandra to test IDG model.

\section{Summary}
In this draft, we calculate the deflection angle of photons that graze the sun within semiclassical IDG model. We find that the deflection angle only depends on $\Lambda/E$. When $\Lambda/E \rightarrow \infty$, $\theta \rightarrow \theta_E$. In other words, we recover the prediction of GR for low energy photons. $\theta \rightarrow 0$ when $\Lambda/E \rightarrow 0$, which means that there is no deflection for sufficiently high energy photons. The transition occurs at range $10^{4}< E/\Lambda < 10^{7}$.

The deviation of deflection angle from $\theta_E$ occurs at X-ray and gamma ray range because $\Lambda>0.01~eV$. It can be tested by X-ray or gamma ray telescopes with good enough angular resolution. However, only Chandra can be possibly used to test this effect. It is interesting to measure the deflection angle of high energy photons and such measurement has never been done before.

\acknowledgments  We thank Prof.~Yi-Zhong Fan, Prof.~Qiang Yuan and Dr. Yuan-Yuan Chen for helpful discussions and suggestions.
This work was supported in part by the 973 National Major State Basic Research and Development of China (No. 2013CB837000), the National Key Program for Research and Development (2016YFA0400200), the Youth Innovation Promotion Association CAS (Grant No. 2016288) and the Natural Science Foundation of Jiangsu Province (Grant No. BK20151608).
\\

\end{document}